\documentclass[twocolumn]{aastex701}
\usepackage{amsmath,amssymb,bm}
\usepackage{parskip}
\usepackage{comment}
\usepackage{bbding}

\shorttitle{Gravitational Lensing Predictions from Wave Simulations of Fuzzy Dark Matter}
\shortauthors{Zhou et al.}

\begin{document}

\title{Gravitational Lensing Predictions from Wave Simulations of Fuzzy Dark Matter}

\author[0009-0001-6796-5347]{Jiajun Zhou}
\affiliation{School of Physics and Astronomy, Beijing Normal University, Beijing 100875, China}
\email{jjzhou@mail.bnu.edu.cn}

\author[0000-0002-8492-4408]{Zhengxiang Li}
\affiliation{School of Physics and Astronomy, Beijing Normal University, Beijing 100875, China}
\affiliation{Institute for Frontiers in Astronomy and Astrophysics, Beijing Normal University, Beijing
102206, China}
\email[show]{zxli918@bnu.edu.cn}

\author[0000-0003-1276-1248]{Amruth Alfred}
\affiliation{Department of Physics, The University of Hong Kong, Pokfulam
Road, Hong Kong}
\affiliation{The Hong Kong Institute for Astronomy and Astrophysics,\\ 
The University of Hong Kong, Pokfulam Road, Hong Kong, P. R. China}
\email[show]{aalfred@hku.hk}

\author[0009-0009-3255-4132]{Ran Gao}
\affiliation{Department of Astronomy, Tsinghua University, Beijing 100084, China}
\email{gaor@mail.tsinghua.edu.cn}

\author[0000-0003-4220-2404]{Jeremy Lim}
\affiliation{Department of Physics, The University of Hong Kong, Pokfulam
Road, Hong Kong}
\affiliation{The Hong Kong Institute for Astronomy and Astrophysics,\\ 
The University of Hong Kong, Pokfulam Road, Hong Kong, P. R. China}
\email{jjlim@hku.hk}

\author[0000-0002-3567-6743]{Zong-Hong Zhu}
\affiliation{School of Physics and Astronomy, Beijing Normal University, Beijing 100875, China}
\affiliation{Institute for Frontiers in Astronomy and Astrophysics, Beijing Normal University, Beijing
102206, China}
\email{zhuzh@bnu.edu.cn}

\begin{abstract}
   \noindent In the cold dark matter paradigm, ultra-light particles are emerging as strong contenders to conventional massive particles. A unique prediction of dark matter comprising such ultra-light particles, known as fuzzy dark matter (FDM), is the presence of strong density modulations throughout galactic halos due to wave interference, which -- when approximated by a Gaussian random field (GRF) -- have been proposed to account for the inability to reproduce the observed positions (when measured at sufficient precisions) and flux ratios of multiply-lensed images of quasars. Here, we predict for the first time the properties of gravitationally lensed images generated from 3-D density fields obtained by wave simulations that directly evolve the Schr\"odinger--Poisson equations. Using a novel framework to project these evolved density fields along various axes of the 3-D halo, we obtain the distribution of perturbations to the positions of lensed images. As an exacting test, we find that particles of mass $10^{-22}$ eV can reproduce the positions of the quadruply-lensed radio jets in system HS 0810+2554 to a level better than that of either the GRF approximation or, to a greater extent, an NFW best-fit solution, both of which rely on accurately capturing the global 3-D density field of dark matter halos. Our work highlights the importance of wave simulations for making accurate FDM lensing predictions and the potential for high-resolution observations of lensed systems to serve as a direct probe of the nature of dark matter.
\end{abstract}


\section{Introduction}
\label{sec:intro}

\noindent The nature of dark matter remains one of the central open questions in modern astrophysics and cosmology. The standard cold dark matter (CDM) framework has been remarkably successful in explaining observations of large-scale structures of the Universe and cosmic microwave background anisotropies \citep{Planck2020}. On galactic and sub-galactic scales, however, the standard CDM paradigm, for which weakly interacting massive particles (WIMPs) are the most theoretically motivated particle candidate, has struggled with several challenges, such as the missing satellite problem~\citep{1999ApJ...516..530K,1999ApJ...522...82K}, the cusp-core problem~\citep{2010AdAst2010E...5D}, and the too-big-to-fail problem~\citep{2011MNRAS.415L..40B,2012MNRAS.422.1203B,Bullock2017}. While these discrepancies remain subject to ongoing debate, an attractive alternative to WIMPs are cold, ultra-light particles which appear to be able to alleviate problems faced by WIMP DM without introducing complex baryonic feedback processes. Ultralight dark matter, initially named as fuzzy cold dark matter~\citep{Hu2000}, is composed of a Bose-Einstein condensate whose de Broglie wavelength can be on the kiloparsec (kpc) scale and comparable to the sizes of galaxies, so that the halo density field is governed by coherent wave dynamics rather than purely classical particle trajectories \citep{Hui2017,Hui2021}. Therefore, ultra-light dark matter is also usually referred to as wave dark matter and abbreviated as $\psi$DM or FDM in the literature. Cosmological and idealized simulations of FDM halos have shown that they naturally develop a soliton-like central core surrounded by an extended halo with granular interference fluctuations \citep{Schive2014,Schwabe2016,Mocz2017,Veltmaat2018}. Due to this unique feature, constraints have been placed on this scenario via several independent observational routes, including Ly$\alpha$ forest~\citep{2017PhRvL.119c1302I,2021PhysRevLett.126.071302,2026arXiv260606969L}, Jeans modeling and dynamics of dwarf galaxies~\citep{2017MNRAS.468.1338C,2020ApJ...893...21S,Broadhurst1,2022PhRvD.106f3517D,Pozo1}, and Milky Way substructure~\citep{2021PhysRevLett.126.091101,2021JCAP...10..043B} (see the reviews by~\cite{Hui2021} and~\cite{2021A&ARv..29....7F} for comprehensive summaries).

Galaxy-scale strong gravitational lensing provides a particularly sensitive probe of the key predictions from FDM models — namely, the small-scale density fluctuations imprinted by wave interference — because small perturbations to the lens potential can produce observable anomalies in image positions, flux ratios, and time delays that are difficult to reproduce with overly smooth macroscopic lens models~\citep{Mao1998,Metcalf2001,Dalal2002,Vegetti2009,Vegetti2010,2025ApJ...991L..27L}. Therefore, perturbations of the gravitational potential in the lens halo due to wave-induced density fluctuations have been widely proposed as a possible origin of astrometric, flux ratio, and time delay anomalies in multiply imaged quasars \citep{Chan2020,2023MNRAS.524L..84P,Amruth2023,2024PhRvD.110h3536L}. In this context, the quadruply lensed quasar HS~0810+2554 is a particularly useful system. Originally identified as a bright quadruply imaged quasar \citep{Reimers2002}, subsequent high-resolution radio observations with e-MERLIN and the European VLBI Network revealed two compact radio components associated with jet activity in the radio-quiet quasar, producing two sets of quadruply lensed radio images \citep{Hartley2019}. These eight radio images with precise astrometric measurements are in significant disagreement with predictions from simple smooth elliptical lens models. Recent work has proposed that FDM perturbations can account for the general level of observed position ($\sim$5-10 mas) and flux-ratio anomalies in lensed quasars, and also specifically in system HS~0810+2554 \citep{Amruth2023}, while another study has argued that increasing model complexity in the form of angular mass components, such as multipole perturbations, and external shear on top of smooth, elliptical power-law profiles may also alleviate position anomalies \citep{Miller2025}. In addition to the question of which physical origin best explains the anomalies, a separate methodological concern remains unresolved. All previous lensing applications of FDM fluctuations have relied on projected perturbation fields generated via Gaussian-random-field (GRF) approximations, rather than on direct three-dimensional Schr\"odinger--Poisson evolution. While such approaches are computationally efficient, it remains an open question whether the lensing signatures they produce are statistically consistent with those obtained from full wave evolution. 

In this work, we first simulate a considerable number of realizations for a wave dark matter halo, whose mass is the same as that of the lens galaxy in HS~0810+2554, directly from evolving the Schr\"odinger--Poisson equations. Next, we develop a three-dimensional framework that projects evolved FDM halos along optimized viewing directions and compares the resulting lensing configurations with the observed relative image geometry. We show that for $m_{\psi}=10^{-22}\,{\rm eV}$, the evolved halos reduce the median position anomaly $A$ of the two radio components to $12$ and $6\,{\rm mas}$, respectively, with some realizations lying within the $\sim 3\sigma$ astrometric uncertainty level, i.e., $A < 3\,{\rm mas}$ \citep{Hartley2019}. This demonstrates that direct Schr\"odinger--Poisson evolution predicts astrometric anomalies consistent with observations at a level comparable to or better than GRF-based approaches, while naturally capturing non-Gaussian statistics that GRF models cannot reproduce. These results establish direct wave evolution as a viable and more physically grounded framework for testing FDM predictions with strong lensing systems.

\section{Simulating Wave Dark Matter Halos}\label{sec:method}

\begin{figure*}[t]
    \centering
    \includegraphics[width=0.98\textwidth]{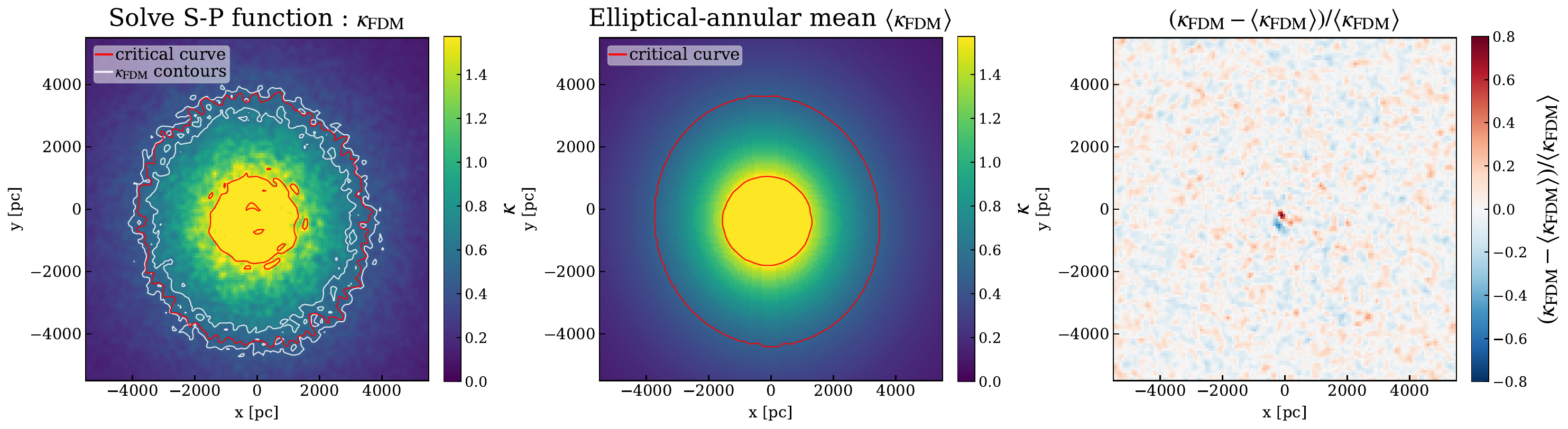}
    \caption{Projected density field of an evolved fuzzy dark matter halo obtained from the Schr\"odinger--Poisson simulation ($m_{\psi}=10^{-22}\,\mathrm{eV}$). From left to right, the panels show the convergence $\kappa_{\rm FDM}$ (projected surface mass density relative to the critical density), its elliptical-annular mean $\langle\kappa_{\rm FDM}\rangle$, and the relative residual between the total and mean convergence $(\kappa_{\rm FDM}-\langle\kappa_{\rm FDM}\rangle)/ \langle\kappa_{\rm FDM}\rangle$. The ellipticity and position angle of the annuli are inferred from the projected convergence distribution. The white contours correspond to $\kappa_{\rm FDM} \sim 0.45$ and $0.60$, respectively, and highlight the density fluctuations around the critical curve, which is shown in red. For visualization, all maps are clipped at their respective 95th-percentile values. See Section~\ref{app:residual_maps} for further discussion.
    }
    \label{fig:map_fdm}
\end{figure*}

\noindent The dynamical evolution of FDM can be simulated using several numerical approaches, including global Fourier pseudospectral methods, finite-difference schemes, local pseudospectral methods, smoothed-particle hydrodynamics, finite-volume solvers, eigenmode expansions, collisionless $N$-body approximations, and so on. In this work, we adopt a global Fourier pseudospectral method, which provides spectral accuracy for the smooth wave fields and effectively suppresses numerical diffusion. This choice is particularly well suited to galaxy lens systems, where the halo mass is of order $10^{11}\, M_\odot$. For a commonly adopted FDM particle mass $m_{\psi}=10^{-22}\,{\rm eV}$, the characteristic de Broglie wavelength is of order $100\,{\rm pc}$, estimated as
\begin{equation}
\lambda_{\mathrm{dB}}=150\left(\frac{10^{-22}\,{\rm eV}}{m_{\psi}}\right)\left(\frac{M_{\mathrm{h}}}{10^{12}\,M_{\odot}}\right)^{-1/3}{\rm pc},
\label{lambda}
\end{equation}
where $M_{\mathrm{h}}$ denotes the total halo mass \citep{Amruth2023,2014PhRvL.113z1302S}. Therefore, the simulations must resolve sub-kiloparsec wave-interference structures with high resolution to capture their lensing effects with sufficient accuracy. For details on the simulation procedure, we refer the reader to the Appendix.

Taking into account the computational cost, the resolution requirements discussed above, and the halo-mass scale relevant to the HS~0810+2554 lens system, we adopt a simulation setup with a grid size of $N=512$, a comoving box size of $L=40~{\rm kpc}$, a total evolution time of $t_{\rm tot}\approx3.3~{\rm Gyr}$ (this sufficiently long timescale allows the evolution to settle into a steady state, and remains within the cosmic age at the lens redshift), FDM particle masses of either $m_{\psi}=10^{-22}~{\rm eV}$ or $10^{-23}~{\rm eV}$, and a total dark matter mass of $M=4\times10^{11}~M_\odot$. The initial condition is constructed from five randomly distributed Gaussian wave packets. These parameter choices are determined only by the approximate halo-mass scale of the lens system with the aim of reproducing the correct Einstein radius and the observed lensing configuration (See Appendix for further discussion).

\begin{figure}
    \centering
 	\includegraphics[width=.9\linewidth]{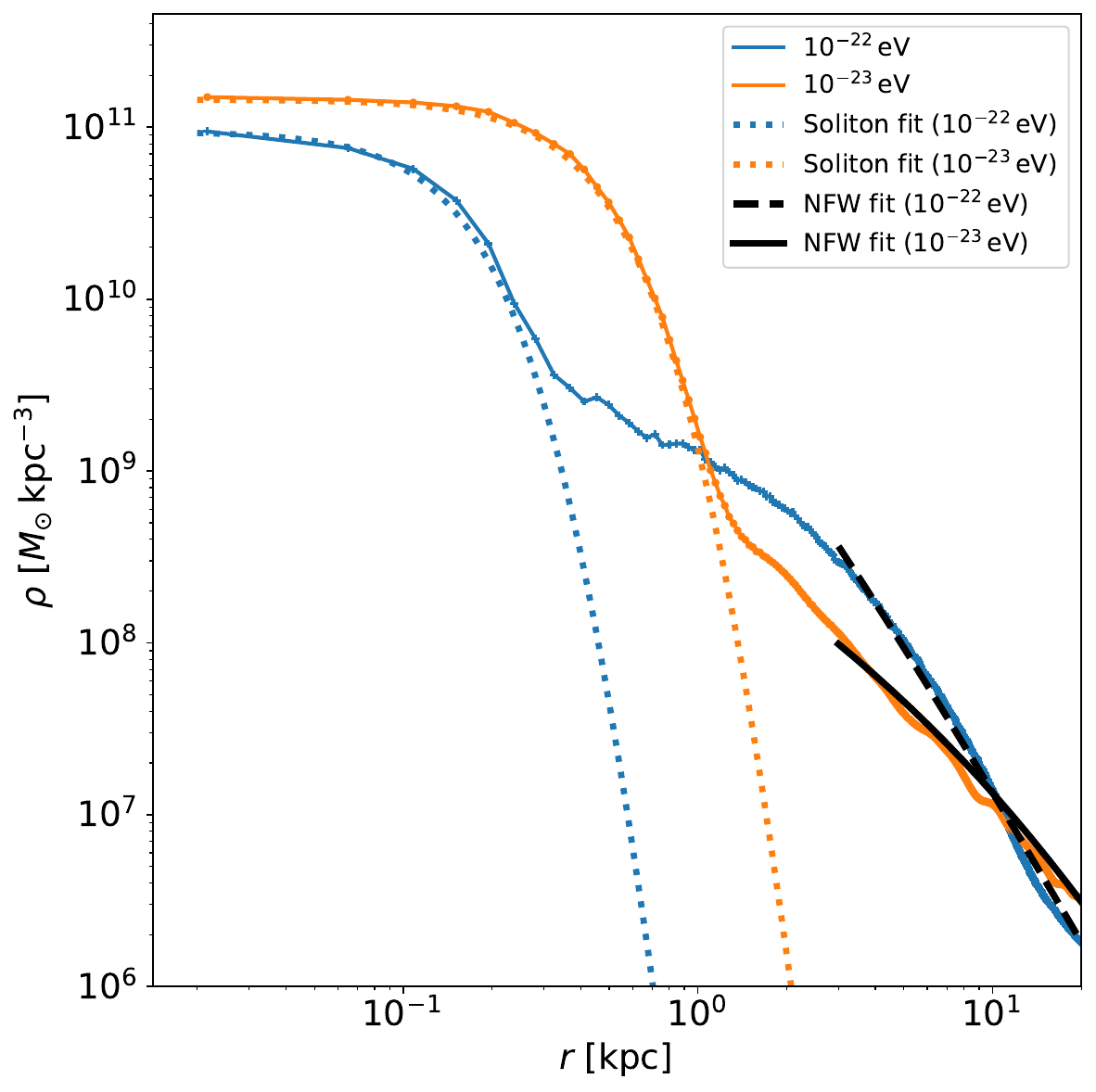}
    \caption{
    Spherically averaged radial density profiles of the evolved fuzzy dark matter halos. The blue and orange curves correspond to particle masses of $m_{\psi}=10^{-22}\,\mathrm{eV}$ and $m_{\psi}=10^{-23}\,\mathrm{eV}$, respectively. The blue and orange dotted curves show the corresponding best-fitting soliton profiles in the inner halo regions \citep{2014PhRvL.113z1302S}. The black dashed and solid curves show the best-fitting NFW profiles for the $m_{\psi}=10^{-22}\,\mathrm{eV}$ and $m_{\psi}=10^{-23}\,\mathrm{eV}$ halos, respectively, obtained using only the radial data beyond the effective Einstein radius.
    }
    \label{fig:rho}
\end{figure}

We generate 1000 independent initial conditions, each containing five randomly sampled three-dimensional Gaussian-wave-packet centers. Evolving these wave-packets for the prescribed period of time with the Schr\"odinger--Poisson solver described in the Appendix, we obtain 1000 three-dimensional FDM halos. As a consistency check, we plot the projected mass density (details in Appendix) and also compute their averaged radial density profiles. Figure~\ref{fig:map_fdm} shows an exemplar realization and its projected mass density (projected convergence $\kappa$), elliptically-averaged radial mean density, and the residual map of the pervasive density fluctuations. It is clear that the density fluctuations produce ubiquitous perturbations to the critical curve (a region of high lensing magnification where lensed images typically form) - we will present exactly how these perturbations affect the properties of lensed images in the next section. The residual projected mass density map illustrates the over- and under-dense fluctuations throughout the halo. In addition, the physical size of the fluctuations can be seen to follow a characteristic scale set by the de Broglie wavelength (which depends on the particle mass and total halo mass), along with a relative fluctuation amplitude of $\sim$10\% in projected density, which is the level expected in galaxy-scale halos \citep{Amruth2023} (see Appendix for detailed relative fluctuation analyses). 

Figure~\ref{fig:rho} shows the radial density profiles for an example simulated halo for two different particle masses with the same initial conditions: $m_{\psi}=10^{-22}~{\rm eV}$ and $m_{\psi}=10^{-23}~{\rm eV}$. The evolved halo exhibits a clear inner--outer transition and a dense central soliton-like core, consistent with the expected structure of an FDM halo from previous simulations \citep{Schive2014,Mocz2017,2025PhRvL.135f1002L}. In the outer regions of the halo, the density slope is similar to that of an NFW profile, again consistent with findings from previous work in the literature \citep{Herrera_Mart_n_2019} and confirms that our numerical evolution produces physically sound halo realizations for the subsequent lensing analysis. The difference in the size of the solitonic core (and hence the de Broglie wavelength) between the two curves illustrates the sensitivity of the evolved halo structure to the FDM particle mass. Notice that the core--halo mass relation exhibits substantial intrinsic scatter and should therefore not be interpreted as a unique or universal relation applicable to individual halos \citep{2017MNRAS.471.4559M,2022MNRAS.511..943C,2022PhRvD.105b3512Y}. In the present work, the relation is shown only as an illustrative reference for the radial-density profile, indicating that the dynamically evolved halo contains a solitonic central component.

We also develop a novel forward-modelling framework to generate lensed images from the three-dimensional density field evolved from the Schr\"odinger--Poisson equations and then compare with the observed image positions for any given lens system. The method involves using pairwise-distance invariants and Procrustes alignment \citep{Schonemann1966} to search for the optimal spatial orientation of the simulated density field and the optimal source position by minimizing the relative geometrical mismatch between the simulated and observed four-image configurations (the search strategy is schematically illustrated in Figure~\ref{fig:method} in the Appendix). The subsequent lensing calculations are performed with the open-source package \texttt{lenstronomy} \citep{Birrer2018,Birrer2021}. For each simulated density field, we determine its principal axis, sample effective viewing orientations, project the rotated density field into a convergence map, and solve the lens equation for a grid of source positions. We again refer the reader to the Appendix for additional details on this procedure.

\section{MULTIPLY-LENSED IMAGES GENERATED BY WAVE-SIMULATED HALOS}
\label{sec:results_discussion}

We now present novel results for the first time on the properties of lensed images generated using the 3-D density fields directly from the full-wave simulations, which are then projected onto 2-D when invoking the conventional thin lens approximation. To date, there have been no studies that have generated and studied lensed images directly using full-wave simulations of FDM, and therefore we also compare our direct predictions with the conventional Gaussian Random Field (GRF) approximation adopted for FDM lensing predictions in previous studies \citep{Chan2020,Laroche_2022,Amruth2023,2026arXiv260116818H}.

\begin{figure}
    \centering

    \begin{minipage}{0.82\linewidth}
        \centering
        \includegraphics[width=\linewidth]{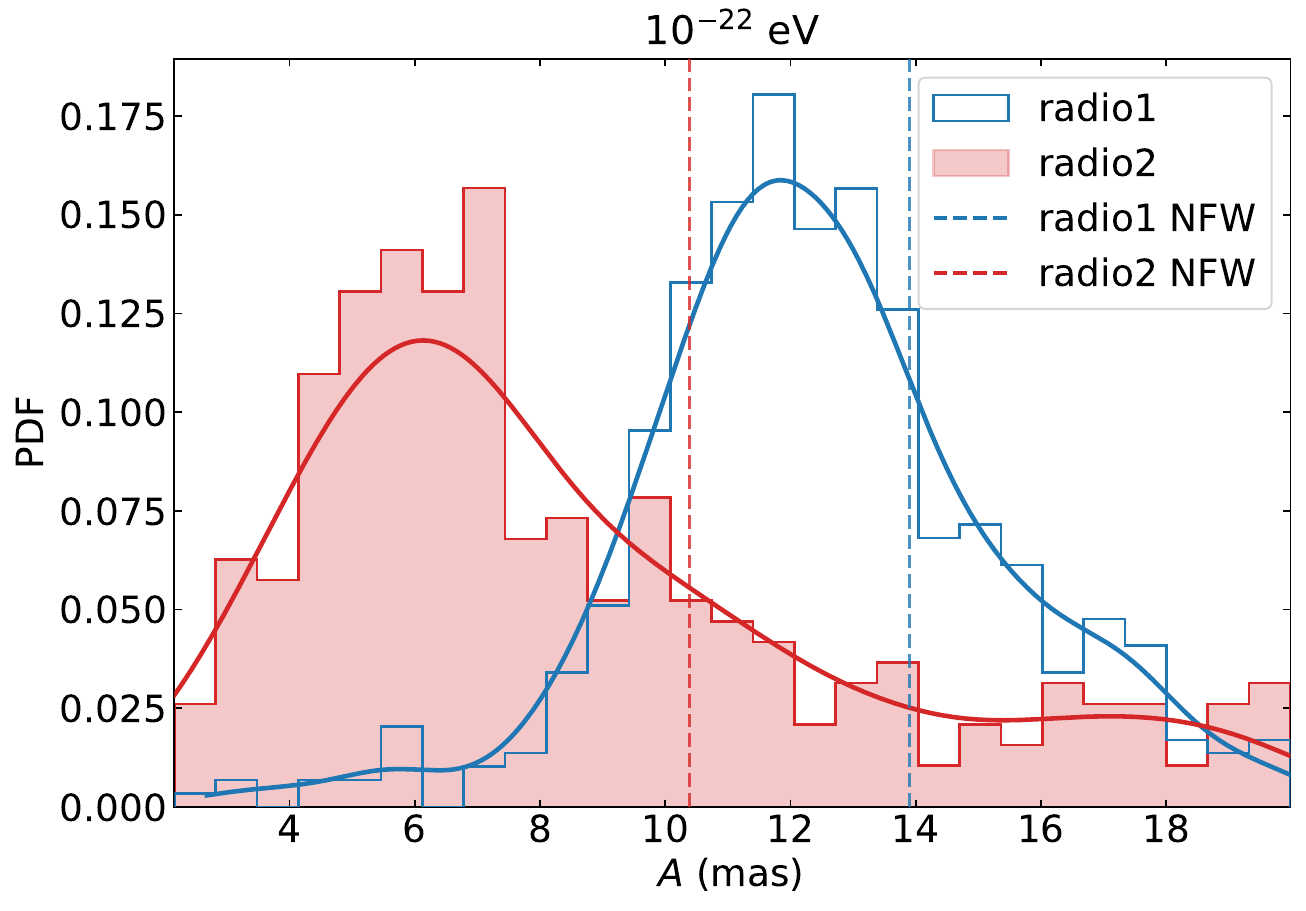}
    \end{minipage}

    \begin{minipage}{0.82\linewidth}
        \centering
        \includegraphics[width=\linewidth]{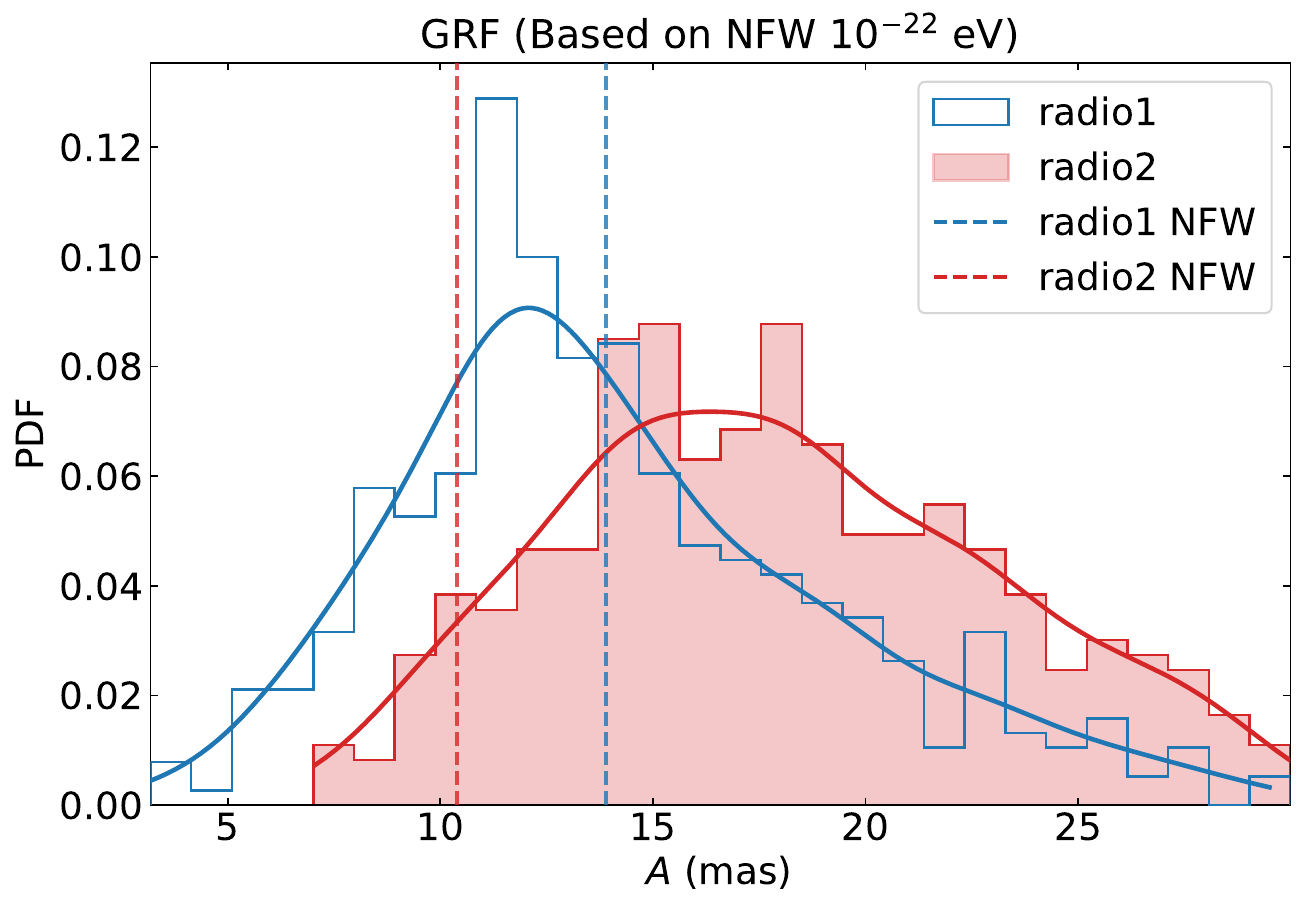}
    \end{minipage}

    \begin{minipage}{0.82\linewidth}
        \centering
        \includegraphics[width=\linewidth]{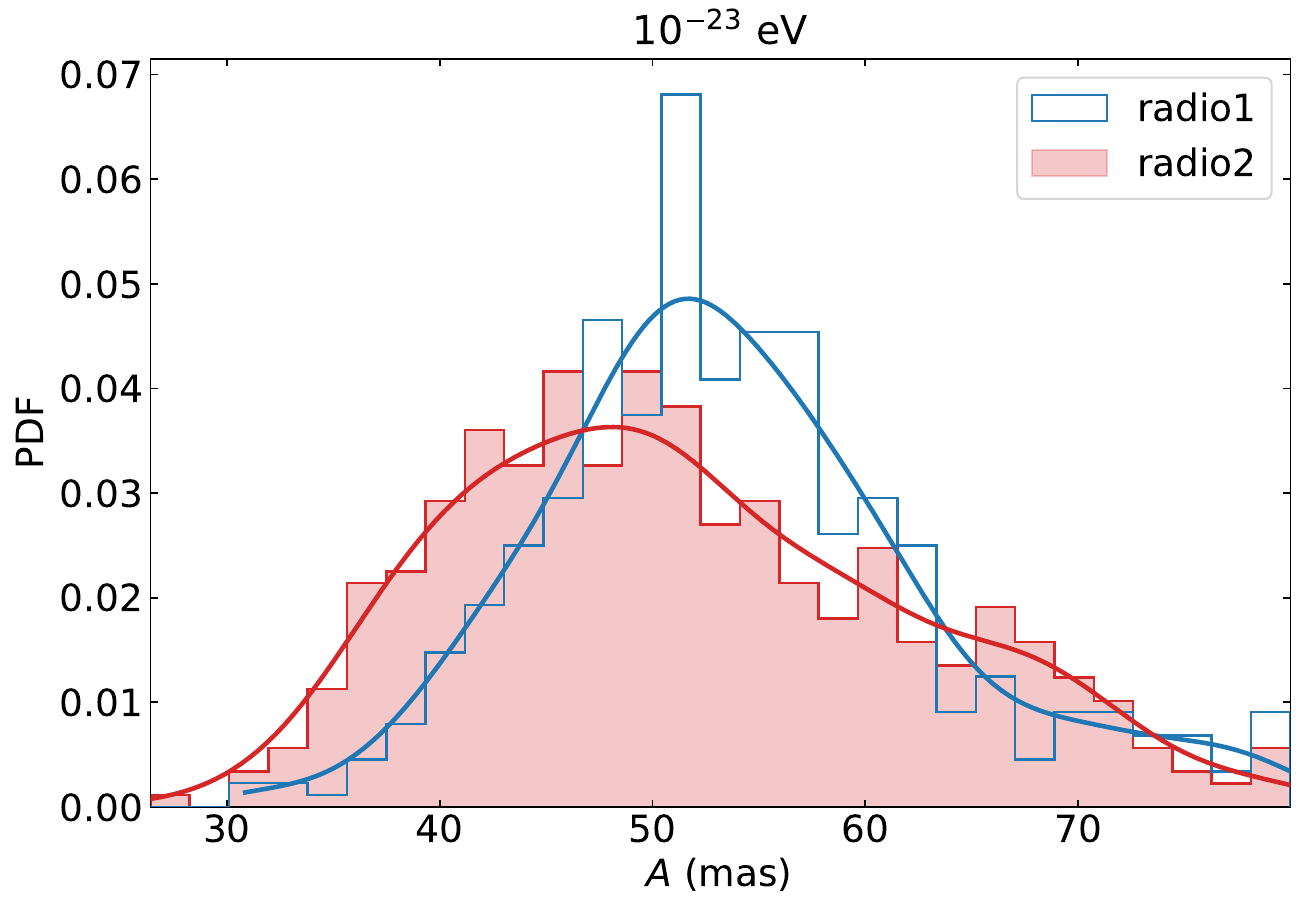}
    \end{minipage}
    \caption{
    Probability distributions of the image-position anomaly $A$.
    (a) Schr\"odinger--Poisson FDM simulations with $m_{\psi}=10^{-22}~{\rm eV}$.
    (b) GRF-based FDM density realizations constructed on top of the best-fitting NFW model.
    (c) Schr\"odinger--Poisson FDM simulations with $m_{\psi}=10^{-23}~{\rm eV}$.
    The blue and red histograms show the normalized distributions for the two radio sources, radio1 and radio2, respectively, while the solid curves denote the corresponding kernel-density-estimation (KDE) fits. The blue and red vertical dashed lines indicate the position anomalies predicted by the best-fitting smooth NFW model for radio1 and radio2, respectively.
    }
    \label{fig:Ar}
\end{figure}

Constructing lensed images for each of the 1000 evolved FDM density fields by forward-lensing a background point source, we quantify the perturbations in the position of the lensed images using the position anomaly, which is the root-mean-square (rms) dispersion (i.e., combining all 4 images) between the predicted and observed image positions, defined as 
\begin{equation}
    A=\left[
    \frac{1}{4}
    \sum_{i=1}^{4}
    \left|
    (\bm{x}_{\mathrm{obs,i}}-\bm{x}_{\mathrm{pred,i}})+(\bm{y}_{\mathrm{obs,i}}-\bm{y}_{\mathrm{pred,i}})
    \right|^2
    \right]^{1/2}.
    \label{eq:A}
\end{equation}
To compare the Schr\"odinger--Poisson evolution with the commonly used GRF approximation, we also generate 1000 GRF realizations. The best-fitting smooth NFW parameters are adopted from \citet{2024PhRvD.110h3536L}, while the GRF perturbations are generated following \citet{Amruth2023}. The resulting probability distribution histograms for all 1000 realizations are shown in Figure~\ref{fig:Ar} for three different cases: $m_{\psi}=10^{-22}~{\rm eV}$, the corresponding predictions for the Gaussian random field approximation and finally for $m_{\psi}=10^{-23}~{\rm eV}$. We separately show the position anomaly for each of the two quadruply-lensed radio sources in the HS 0810+2554 system. 

It is immediately apparent that there is a wide distribution in the position anomaly relative to the best-fit smooth lens model. For the wave-simulated $m_{\psi}=10^{-22}~{\rm eV}$ case, the median position anomaly is $\sim 12~{\rm mas}$ and $\sim 6~{\rm mas}$ for the two radio sources, both indicating better agreement with the observed image positions than the best-fit NFW lens model predictions. There also exist some realizations where the position anomaly is within $\sim 3\sigma$ of the measurement uncertainties in positions for both radio sources, demonstrating that the wave-simulated halos can indeed reproduce the observed image positions. In comparison, the position anomaly distribution for the lensed images constructed using the GRF approximation is shifted to larger values, especially for one of the radio components, and rarely falls below the $5\sigma$ level. In general, the GRF approximation produces larger fluctuations in the image positions and worse positional agreement relative to the wave-simulated halos. This comparison demonstrates that wave-simulated halos may more efficiently reproduce the observed image positions than the GRF-based realizations. When we look at the predictions for the lighter particle mass $m_{\psi}=10^{-23}~{\rm eV}$, it produces, as expected, substantially larger fluctuations in the image positions, with a median position anomaly of $\sim$50 mas which is a factor of $\sim$4 larger than those produced by $m_{\psi}=10^{-22}~{\rm eV}$.

\begin{figure*}[t]
    \centering
    \includegraphics[width=0.98\textwidth]{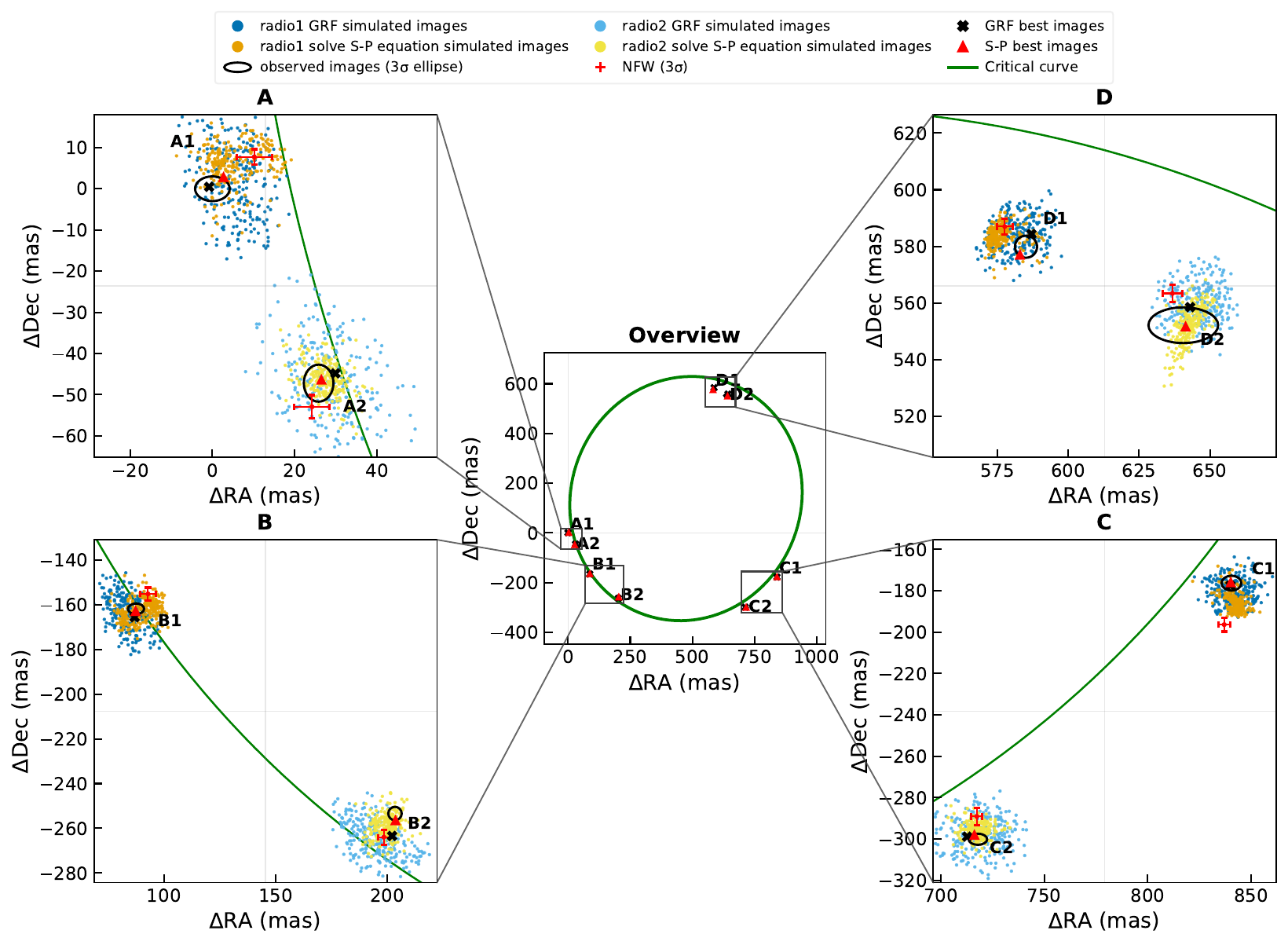}
    \caption{
    Predicted and observed lensed image positions for the two radio components of HS~0810+2554. The central panel shows the full image configuration, while the surrounding panels zoom into the individual images. To avoid visual overcrowding, only the 300 best-matching realizations with the smallest position anomaly $A$ are shown, with the images denoted by colored points. The black solid ellipses denote the $3\sigma$ measurement uncertainties of the image positions. The NFW-predicted image positions and their corresponding $3\sigma$ uncertainties obtained from MCMC simulations, adopted from~\citet{2024PhRvD.110h3536L}, are shown as red + symbols. The black $\times$ symbols denote the best-matching GRF image positions, while the red $\blacktriangle$ symbols denote the best-matching S--P simulated image positions. The green curve denotes the critical curve of the best-fitting elliptical NFW model for HS~0810+2554. }
    \label{fig:image_scatter}
\end{figure*}

Now that we have made a statistical prediction for the distribution of position anomalies averaged over all four lensed images (see Equation~\ref {eq:A}), we proceed to address the more stringent question of whether the positions of each lensed image can be reproduced. Figure~\ref{fig:image_scatter} shows the predicted image positions for both radio components for 300 unique FDM halos (as plotting the images for all 1000 halos makes visual comparison difficult). It is readily apparent that the Schr\"odinger--Poisson and GRF approaches produce an image `cloud' that overlaps with the observed image positions (indicated with 3$\sigma$ error ellipses) for all eight lensed images. Upon closer inspection, the wave-simulated halos (orange and yellow points) tend to produce more compact image clouds around the observed positions relative to the GRF predictions (blue and teal points). The best-fit NFW predictions (red and blue crosses) are in significantly higher tension ($\geq5\sigma$) with the observed image positions for all images except one. We remind the reader that since the de Broglie wavelength (and therefore physical size of fluctuations) is smaller than the separation distances between the observed images, the image predictions for different realizations are independent and can in theory be combined to construct a single realization that can successfully reproduce all eight lensed images as long as there exist realizations where the predicted images can reproduce the observed images for each of the eight lensed images.

It is remarkable that the wave-simulated halos can reproduce the observed image positions of HS~0810+2554 to within 3$\sigma$, despite involving no best-fitting procedure: the halos are evolved from initial conditions without any control over their final states. This is in sharp contrast to the GRF and NFW approaches, which rely on best-fit models obtained through conventional minimization techniques typically used in lens modelling.

\section{SUMMARY}\label{sec:conclusions}

In this work, we report the first predictions for the properties of gravitationally lensed images produced by FDM halos evolved directly from the Schr\"odinger--Poisson equations, with a focus on the perturbations to the lensed image positions due to the pervasive density fluctuations present throughout the entire FDM halo. As a specific test of our novel predictions, we assess whether the positions of the well-studied quadruply-lensed quasar HS~0810+2554, for which very high angular resolution VLBI observations reveal two compact radio sources, can be successfully reproduced with FDM halos. To this end, we developed a forward-modelling framework that connects three-dimensional FDM density fields obtained from Schr\"odinger--Poisson evolution with the generation of multiply-lensed image configurations, where for each simulated density field we projected the halo along a set of effective viewing directions and searched for the source position that best reproduces the observed relative four-image geometry.

We find that the lensing configurations generated from wave-simulated halos with Schr\"odinger--Poisson evolution for a particle mass of $m_{\psi}=10^{-22}$eV can reproduce the image positions of the two radio components to within $\sim$3 mas (a 3$\sigma$ discrepancy) in the best realization, but still better than the calculation with the Gaussian random field approximation which leaves behind best-fit position anomalies of $\sim$7 mas for one of the radio components and the best-fit NFW prediction which leaves behind the most severe position anomalies of $\sim$12 mas. The wave-simulated halos also produce a more compact distribution in position anomalies relative to the GRF approximation as can be seen by the distributions in Figure~\ref{fig:Ar}. Repeating the analysis for a lighter particle mass of $m_{\psi}=10^{-23}~{\rm eV}$ results in substantially larger position anomalies, as expected due to the larger de Broglie wavelength and therefore larger physical scale of the density fluctuations. Given the sensitivity of the position anomaly to the FDM particle mass, future high angular resolution observations of lensed systems will provide a promising route for narrowing down the allowed mass range for the DM particle. From the differences uncovered between the wave-simulated and GRF halos, it is evident that full wave evolution may be necessary to predict gravitational lensing phenomena at sufficient precision to make inferences on the nature of the dark matter particle using existing and future observations of lensed systems. At the same time, the annular residual statistics and Gaussianity tests (see Appendix for details) support the notion that the GRF approach may be a useful statistical approximation of the projected density field for preliminary calculations.

To better understand the differences between the S--P and GRF lensing predictions, we further compare their projected density statistics in the Appendix. Although the two approaches show broadly similar fluctuation amplitudes and local Gaussian statistics, these similarities do not imply identical lensing predictions, since lensing depends on the full projected mass distribution rather than on the fluctuations alone. The GRF construction relies on a prescribed smooth halo and a Gaussian statistical description, whereas the S--P approach self-consistently evolves the complete three-dimensional density field and retains additional spatial correlations, mode coupling, and non-Gaussian higher-order structure. The improved GRF prediction obtained when its smooth component and fluctuation amplitude are derived from the S--P density field further illustrates this point. Thus, GRF remains an efficient approximation at moderate precision, while direct S--P evolution becomes increasingly important for high-precision lensing predictions. For details on the origin of the improved S--P lensing predictions relative to the GRF approximation, we refer the reader to Appendix~\ref{app:sp_grf_lensing_difference}.

This work lays the groundwork for forthcoming tests of the nature of dark matter using gravitationally lensed systems. One particularly important direction is the longstanding problem of flux-ratio anomalies in multiply lensed quasar \citep{Mao1998,Metcalf_2002,Keeton_2003,Mao_2004,wyn,Kochanek_2004,Xu_2015,Shajib_2018}. Previous studies have suggested that the granular density fluctuations produced by FDM may provide a simple and physically credible explanation for such anomalies, but a systematic analysis based on fully evolved Schr\"odinger--Poisson wave simulations has not yet been carried out. In follow-up work, we plan to extend the present framework beyond image positions to investigate the predicted image magnifications and flux ratios, time delays, and resolved image morphologies, as well as their interplay with density projection axes and the influence of baryonic components in the lens model.
Because full three-dimensional wave simulations are computationally expensive, future work will examine larger simulation boxes with more efficient approaches, such as Schwarzschild methods \citep{2022PhRvD.105b3512Y}, while ensuring sufficient accuracy in the strong-lensing region.

\section{Acknowledgements}
We would like to thank Jianxiang Liu, Kai Liao, Liang Dai, and Shuxun Tian for their helpful discussions. We also thank the editor and the referee for their helpful comments and suggestions, which improved the presentation of this paper. This work was supported by the National Key Research and Development Program of China Grant Nos. 2023YFC2206702, and 2021YFC2203001; National Natural Science Foundation of China under Grants Nos.11920101003, 12021003, 11633001, 12322301, and 12275021; the Strategic Priority Research Program of the Chinese Academy of Sciences, Grant No. XDB2300000 and the Interdiscipline Research Funds of Beijing Normal University. A.A. acknowledges support from the Research Grants Council Junior Research Fellowship Scheme. A.A. and J.L. acknowledge support from the Research Grants Council of Hong Kong under the General Research Fund grant No. 17304425, and also acknowledge support from the Seed Fund for Collaborative Research from the University of Hong Kong. 

\newpage 

\section{Appendix}

\subsection{Pseudospectral Wave Simulation Prescription}
\label{app:simulation}
In the non-relativistic limit, the FDM field is governed by the Schr\"odinger--Poisson equations \citep{Hui2017,2025PhRvD.111h1302G},
\begin{equation}
\begin{aligned}
i\hbar \frac{\partial \psi}{\partial t}
&=\left[-\frac{\hbar^2}{2m}\nabla^2+m\Phi\right]\psi, \\
\nabla^2 \Phi
&=4\pi G M\left(|\psi|^2-\overline{|\psi|^2}\right),
\end{aligned}
\label{eq:schrodinger_poisson}
\end{equation}
where $\psi$ is the complex wave function, $\Phi$ is the Newtonian gravitational potential, $m$ is the FDM particle mass, and $M$ is the total dark matter mass in the computational box. We normalize the wave function according to
\begin{equation}
    \int_{\rm box} |\psi(t,\mathbf{r})|^2\,{\rm d}^3\mathbf{r}=1,
\end{equation}
such that $\rho=M|\psi|^2$. Consequently,
\begin{equation}
    \overline{|\psi|^2}
    =
    V^{-1}
    \int_{\rm box}|\psi(t,\mathbf{r})|^2\,{\rm d}^3\mathbf{r}
    =
    V^{-1},
\end{equation}
with $V$ denoting the box volume. The subtraction of $\overline{|\psi|^2}$ removes the homogeneous density mode in the periodic Poisson solver.

The spatial resolution is chosen to resolve both the FDM de Broglie wavelength and the physical scale corresponding to the observed image-position anomaly. For a cubic box of side length $L$ and grid size $N^3$, the grid spacing is
\begin{equation}
    \Delta x=\frac{L}{N}.
\end{equation}
We require
\begin{equation}
    \Delta x < \lambda_{\rm dB},
\end{equation}
and, more specifically, several grid cells per de Broglie wavelength are used to resolve the interference pattern of the FDM wave field. In addition, an angular anomaly $\Delta\theta$ at the lens redshift corresponds to a physical lens-plane scale
\begin{equation}
    \Delta l
    =
    D_{\rm l}\Delta\theta ,
\end{equation}
where $D_{\rm l}$ is the angular-diameter distance to the lens. The grid spacing is therefore also required to be sufficiently smaller than the physical scale associated with the observed tens-of-milliarcsecond image-position offsets.

The time evolution is performed using a split-step pseudo-spectral method to reduce the leading time-integration error. Over one time step $\Delta t$, the potential and kinetic operators are applied separately \citep{1984JCoPh55203T,2026LRCA121S},
\begin{equation}
\begin{aligned}
&\psi(\boldsymbol{r}, t+\Delta t)\\
& \approx \mathcal{K}(t+\Delta t, \Delta t / 2) \mathcal{D}(t+\Delta t / 2, \Delta t) \mathcal{K}(t, \Delta t / 2) \psi(\boldsymbol{r}, t)\\&+\mathcal{O}\left(\Delta t^3\right),
\end{aligned}
\end{equation}
where
\begin{equation}
\begin{aligned}
& \mathcal{K}(t, \Delta t)=\exp \left[-i \frac{m}{\hbar} \Delta t \Phi(t)\right], \\
& \mathcal{D}(t, \Delta t)=\exp \left[i \frac{\hbar}{2 m} \Delta t \nabla^2\right].
\end{aligned}
\end{equation}

Accordingly, the time step is selected to resolve the fastest phase oscillations in both the kinetic and potential evolution. We impose
\begin{equation}
    \Delta t
    \lesssim
    C_t
    \min
    \left[
     \frac{2 \pi \hbar}{m} \frac{1}{|\Phi|_{\rm max}},
     \frac{4 m}{\pi \hbar} \Delta x^2
    \right],  \label{eq:time_step_constraint}
\end{equation}
where $C_t<1$ is a safety factor, $|\Phi|_{\rm max}$ is the maximum gravitational potential amplitude. This condition is used as a phase-resolution criterion for avoiding phase aliasing in the split-step evolution.

The total integration time is constrained by the cosmological age of the lens system. The source and lens redshifts of HS~0810+2554 are taken to be
\begin{equation}
    z_{\rm s}=1.51,
    \qquad
    z_{\rm l}=0.89 .
\end{equation}
The cosmic age at redshift $z$ is
\begin{equation}
    t(z)
    =
    \int_z^{\infty}
    \frac{{\rm d}z'}
    {(1+z')H(z')},
\end{equation}
with
\begin{equation}
    H(z)
    =
    H_0
    \left[
    \Omega_{\rm m}(1+z)^3
    +
    \Omega_\Lambda
    \right]^{1/2}.
\end{equation}
The FDM density field used for lensing is required to correspond to the lens epoch, and hence the total physical evolution time cannot exceed the cosmic age at $z_{\rm l}$. For the adopted Planck 2020 cosmology \citep{Planck2020}, this provides an upper bound of approximately
\begin{equation}
    t(z_{\rm l}=0.89)
    \simeq
    6.3\,{\rm Gyr}.
\end{equation}
This upper limit ensures that the simulated FDM halo remains physically consistent with the cosmic time available before the lensing event.

\subsection{Forward-Modelling Lensed Images}
\label{app:method}

Starting from the simulated density field $\rho(\bm{r}), \bm{r}=(x,y,z)$, we first determine its principal axis. The mass-weighted centroid is computed as
\begin{equation}
    \bar{\bm{r}}
    =
    \frac{\int \rho(\bm{r})\bm{r}\,{\rm d}^3\bm{r}}
    {\int \rho(\bm{r})\,{\rm d}^3\bm{r}},
\end{equation}
and the corresponding second-moment tensor is
\begin{equation}
    \mathcal{I}_{ij}
    =
    \int \rho(\bm{r})
    \left(r_i-\bar{r}_i\right)
    \left(r_j-\bar{r}_j\right)
    {\rm d}^3\bm{r}.
\end{equation}
The eigenvector associated with the largest eigenvalue of $\mathcal{I}_{ij}$ is defined as the principal axis of the density distribution and is denoted by $\hat{\bm{e}}_{\rm p}$.

\begin{figure}
    \centering

    \begin{minipage}{0.88\linewidth}
        \centering
        \includegraphics[width=\linewidth]{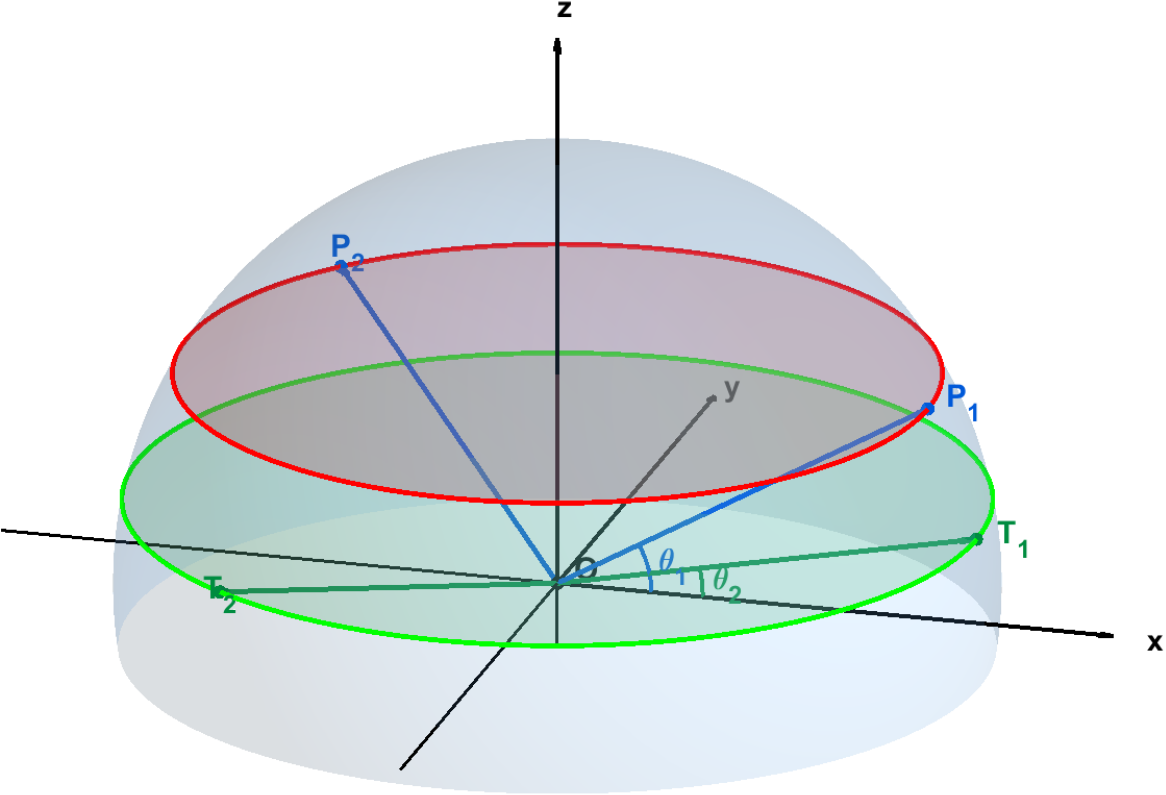}

        \vspace{0.1cm}
        \textbf{(a)}
    \end{minipage}

    \vspace{0.35cm}

    \begin{minipage}{0.72\linewidth}
        \centering
        \includegraphics[width=\linewidth]{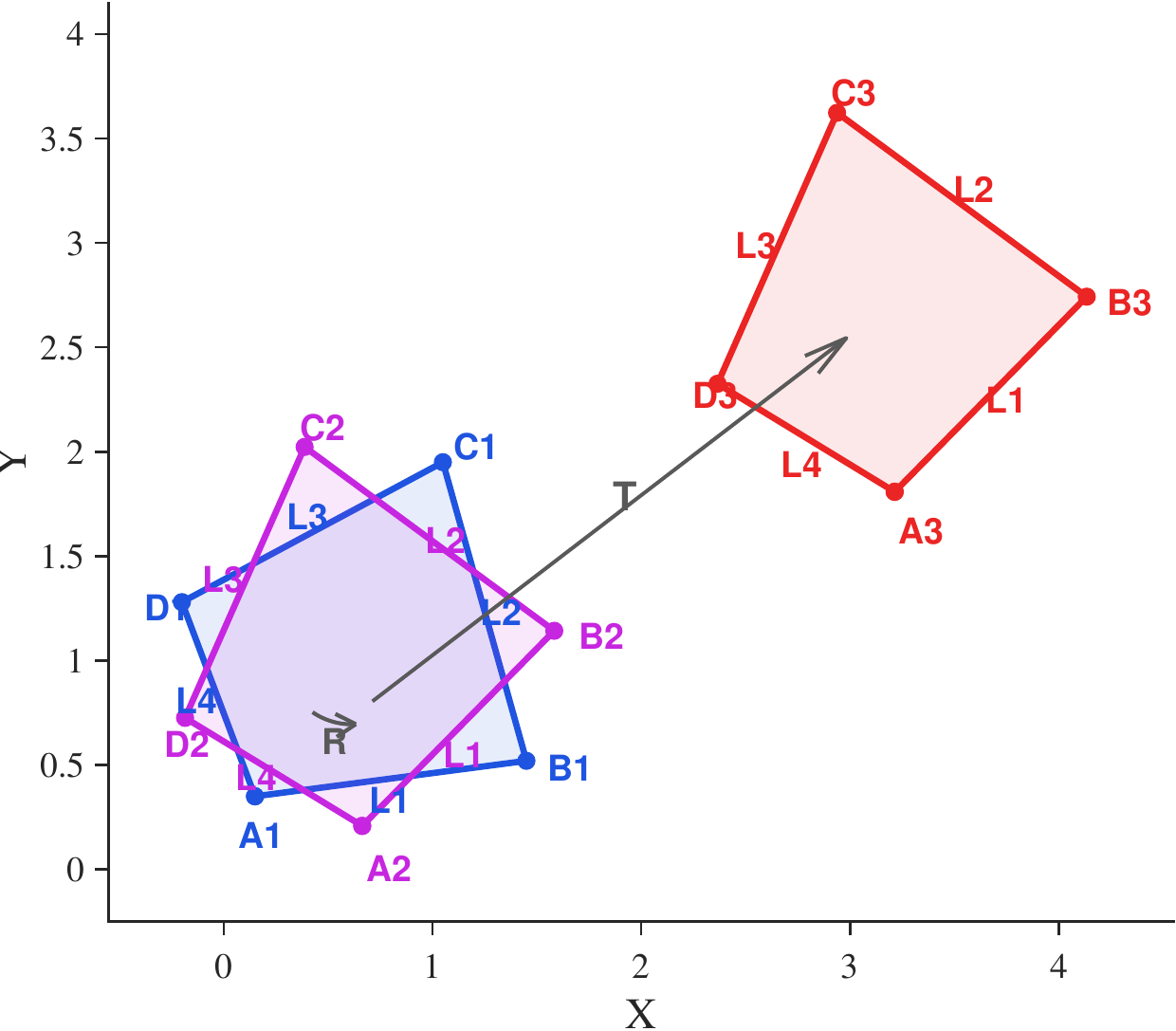}

        \vspace{0.1cm}
        \textbf{(b)}
    \end{minipage}

    \caption{
    Schematic illustration of the reduced matching procedure. (a) The line of sight is fixed along the $z$-axis, while the principal axis of the three-dimensional density field is rotated to representative directions in the positive $xz$ plane. (b) Four-image configurations related by translation, rotation, or reflection are treated as geometrically equivalent. In the actual calculation, all six pairwise distances among the four images are used, including the two diagonal distances not explicitly labelled in the schematic figure. See Section~\ref{app:method} for further discussion.
    }
    \label{fig:method}
\end{figure}

The line of sight is fixed to the $z$-axis. Varying the viewing direction of the density field is therefore implemented equivalently by rotating the density field itself. For each candidate direction $\hat{\bm{e}}_k$, we construct a rotation matrix $\bm{R}_k$ satisfying
\begin{equation}
    \bm{R}_k\hat{\bm{e}}_{\rm p}
    =
    \hat{\bm{e}}_k ,
\end{equation}
and obtain the rotated density field
\begin{equation}
    \rho'_k(\bm{r})
    =
    \rho\left(\bm{R}_k^{-1}\bm{r}\right).
\end{equation}
To avoid redundant sampling of geometrically equivalent orientations, we sample $N_{\rm dir}=10^3$ representative unit vectors in the positive $xz$ plane,
\begin{equation}
    \hat{\bm{e}}_k
    =
    \left(\cos\theta_k,0,\sin\theta_k\right),
    \qquad
    0\leq \theta_k \leq \frac{\pi}{2}.
\end{equation}
This reduced sampling is sufficient because the remaining equivalent configurations differ only by image-plane rotations or mirror transformations, which are fully accounted for in our subsequent image position analysis. As shown in Figure~\ref{fig:method}(a), when the principal axis is rotated to the $P_2$ or $T_2$ direction, the resulting projected surface density is equivalent to an in-plane rotation of that obtained for the corresponding $P_1$ or $T_1$ direction. Thus, the associated image configurations differ only by a two-dimensional rotation. Similarly, reversing the principal axis along the $z$ direction produces a projected surface density that is the mirror counterpart of the positive-$z$ case.

For each rotated density field, the projected surface mass density is obtained under the thin-lens approximation by integrating along the line of sight,
\begin{equation}
    \Sigma_k(x,y)
    =
    \int \rho'_k(x,y,z)\,{\rm d}z,
\end{equation}
and the corresponding convergence is
\begin{equation}
    \kappa_k(x,y)
    =
    \frac{\Sigma_k(x,y)}{\Sigma_{\rm crit}} .
\end{equation}
We then uniformly sample $N_{\rm src}=10^4$ source positions $\bm{\beta}_m=(\beta_{x,m},\beta_{y,m})$ on the source plane. For each source position, the lens equation
\begin{equation}
    \bm{\beta}_m
    =
    \bm{\theta}
    -
    \bm{\alpha}_k(\bm{\theta}),
\end{equation}
is solved with package \textit{lenstronomy}, yielding the simulated image positions on the image plane.

For a four-point configuration $\{P_i\}_{i=1}^4$, the pairwise distances form a symmetric matrix $\boldsymbol{D}$ with $D_{ij}=|P_i-P_j|$ and $D_{ii}=0$, yielding $\binom{4}{2}=6$ independent distances. After removing the three degrees of freedom associated with planar rigid motions, the configuration has five intrinsic degrees of freedom. Thus, the six pairwise distances constitute an overdetermined but complete set of invariants for characterizing the relative four-image geometry. The six pairwise distances are sorted in ascending order for both the observed and simulated image configurations. We then compute the resulting six-dimensional distance residual vector and use its $L_2$ norm as the geometry-mismatch metric. For each principal-axis direction, the source position that minimizes this norm is selected as the optimal source position.

After this source selection, we further quantify the image-position anomaly using Procrustes alignment. The simulated and observed image-position matrices are denoted by $X,Y\in\mathbb{R}^{4\times2}$. After subtracting their centroids, we define
\begin{equation}
    H=X_{\rm c}^{\rm T}Y_{\rm c}.
\end{equation}
With the singular value decomposition (SVD)
\begin{equation}
    H=U\Lambda V^{\rm T},
\end{equation}
the optimal orthogonal transformation is
\begin{equation}
    R=UV^{\rm T},
    \qquad R\in O(2),
\end{equation}
where both rotations and reflections are allowed. The translation vector is
\begin{equation}
    T=\bar{\bm{y}}-\bar{\bm{x}}R,
\end{equation}
and the optimally aligned simulated image positions are
\begin{equation}
    X'=XR+T .
\end{equation}
We define the image-position anomaly as
\begin{equation}
    A=\left[
    \frac{1}{4}
    \sum_{i=1}^{4}
    \left|
    \bm{x}'_i-\bm{y}_i
    \right|^2
    \right]^{1/2},
\end{equation}
where $\bm{x}'_i$ and $\bm{y}_i$ denote the positions of the i-th aligned simulated and observed image, respectively. The above procedure determines the optimal source position for a given principal-axis direction. By comparing the residual anomaly $A$ over all sampled directions, we identify the best-matching pair $(\hat{\bm{e}}_{k_{\rm best}}, \bm{\beta}_{m_{\rm best}})$, which specifies the preferred spatial orientation of the simulated FDM density field and the source position that most closely reproduce the observed relative image geometry.

As an example to demonstrate the superiority of this approach, the GRF-based $A$ distribution obtained with our image-matching procedure is narrower and reaches a smaller minimum $A$ than those derived from conventional nearest-neighbor image matching methods which evaluate the residual in a fixed coordinate frame. Thus, a global translation or rotation between the simulated and observed configurations is counted as part of the anomaly, even when the two systems have the same intrinsic relative geometry. This is important for projected three-dimensional FDM density fields, whose absolute image-plane origin and orientation are not fixed a priori by the observations. Since the observed image positions are relative, global translations and rotations of the simulated configuration should not be penalized. Neglecting this equivalence can artificially increase the inferred image-position anomaly by conflating coordinate-frame mismatch with genuine geometrical residuals. In our analysis, this ambiguity is removed through our use of pairwise-distance invariants and Procrustes alignment.

\subsection{Comparison with Gaussian Random Field Approximation
}
\label{app:comparison}

We begin by computing the annular statistics of the Schr\"odinger--Poisson (S--P) and Gaussian random field (GRF) density fields as functions of projected radius. Figure~\ref{fig:statistics} shows the annular residual root-mean-square (rms) amplitude and the corresponding fractional residual rms. The two approaches exhibit broadly similar radial trends: the fluctuation amplitude is largest in the inner halo and decreases toward larger projected radii. The S--P and GRF curves become close near the effective Einstein radius, where the strong-lensing observables are most sensitive to local convergence perturbations.

This close correspondence indicates that the GRF construction captures the leading fluctuation scale and low-order projected statistics of the dynamically evolved FDM field on the lensing-relevant scales. Small residual differences are expected from finite-sampling effects and from the self-consistent S--P wave evolution, which can generate spatial correlations, mode coupling, and weak non-Gaussian higher-order structure beyond the variance and correlation scale specified in a GRF; nevertheless, the overall consistency demonstrates that the GRF provides a robust and efficient statistical approximation to the S--P results.

We further test whether the residual fluctuations generated by the S--P evolution are locally consistent with Gaussian statistics. Since the fluctuation variance varies with radius, the Gaussianity test is performed within narrow annuli rather than over the full projected field. To match the adopted de Broglie wavelength, $\lambda_{\rm dB}\sim 120~{\rm pc}$, each annulus in Figure~\ref{fig:gaussianity} is assigned a radial width of $\Delta R=120~{\rm pc}$.

The standardized residual distributions of the S--P and GRF realizations closely overlap in the selected annuli and are well described by the standard normal distribution. This is also supported by the normal quantile--quantile plots, where both sets of residuals follow the Gaussian reference line over most of the distribution, with only mild deviations in the tails. The small skewness, near-Gaussian kurtosis, and large Kolmogorov--Smirnov $p$-values further show that the S--P residuals are statistically consistent with Gaussian fluctuations at fixed radius. These results support the use of GRF realizations as an accurate and efficient statistical approximation to S--P-generated FDM density fluctuations, while the full S--P evolution remains essential for resolving the detailed dynamically generated wave interference pattern of individual halo realizations.

\subsection{Residual Convergence Fields in the Schr\"odinger--Poisson and GRF Approaches}
\label{app:residual_maps}

As a supplement to Figure~\ref{fig:map_fdm}, We compare the residual convergence fields obtained from a S--P realization and a GRF realization. We also describe the procedure used to define the smooth reference field and
the corresponding fractional residual in each case.

The S--P calculation directly produces an evolved three-dimensional density field and does not require the introduction of an underlying analytic smooth halo. It is therefore inappropriate to define the S--P fluctuation field by simply subtracting an independently specified NFW profile. To illustrate this point, the projected S--P convergence may be written schematically as
\begin{equation}
    \kappa_{\rm FDM}
    =
    \kappa_{\rm sm}
    +
    \Delta\kappa_{\rm wave},
\end{equation}
where $\kappa_{\rm sm}$ represents the intrinsic smooth component of the evolved halo and $\Delta\kappa_{\rm wave}$ denotes the wave-induced fluctuations. Subtracting an NFW model would give
\begin{equation}
    \kappa_{\rm FDM}-\kappa_{\rm NFW}
    =
    \Delta\kappa_{\rm wave}
    +
    \left(\kappa_{\rm sm}-\kappa_{\rm NFW}\right).
\end{equation}
The resulting quantity would therefore combine the physical wave fluctuations with the mismatch between the intrinsic smooth component and the adopted NFW profile. Such a residual would not provide a clean measurement of the fluctuation field generated by the S--P evolution.

We instead estimate the smooth background directly from the projected S--P convergence map. Because the evolved halo is approximately elliptical in projection, averaging within circular annuli would mix pixels corresponding to different effective radii and could introduce an artificial residual associated with the halo ellipticity. We therefore determine the projected principal axes, axis ratio, and position angle from the mass-weighted second-moment tensor of the projected density distribution. These quantities are then used to define concentric elliptical annuli. The mean convergence within each annulus is assigned to the corresponding elliptical radius, yielding the smooth reference field $\langle\kappa_{\rm FDM}\rangle_{\rm ell}$.

The absolute and fractional S--P residuals are consequently defined as
\begin{equation}
    \Delta\kappa_{\rm SP}
    =
    \kappa_{\rm FDM}
    -
    \langle\kappa_{\rm FDM}\rangle_{\rm ell},
    \label{eq:sp_absolute_residual}
\end{equation}
and
\begin{equation}
    \delta\kappa_{\rm SP}
    =
    \frac{
        \kappa_{\rm FDM}
        -
        \langle\kappa_{\rm FDM}\rangle_{\rm ell}
    }{
        \langle\kappa_{\rm FDM}\rangle_{\rm ell}
    },
    \label{eq:sp_fractional_residual}
\end{equation}
respectively. This construction removes the large-scale elliptical radial profile while retaining nonaxisymmetric structures generated by the wave evolution.

The GRF realization is constructed differently. Following the prescription of \citet{Amruth2023}, a statistically homogeneous fluctuation field with zero ensemble mean is superposed on a prescribed smooth convergence profile. The smooth component is taken to be the NFW model with parameters adopted from \citet{2024PhRvD.110h3536L}. The GRF convergence field may therefore be expressed as
\begin{equation}
    \kappa_{\rm GRF}
    =
    \kappa_{\rm NFW}
    +
    \Delta\kappa_{\rm GRF},
\end{equation}
for which subtraction of the NFW component directly recovers the imposed GRF fluctuation. The corresponding residuals are
\begin{equation}
    \Delta\kappa_{\rm GRF}
    =
    \kappa_{\rm GRF}
    -
    \kappa_{\rm NFW},
    \label{eq:grf_absolute_residual}
\end{equation}
and
\begin{equation}
    \delta\kappa_{\rm GRF}
    =
    \frac{
        \kappa_{\rm GRF}
        -
        \kappa_{\rm NFW}
    }{
        \kappa_{\rm NFW}
    }.
    \label{eq:grf_fractional_residual}
\end{equation}
Thus, although the S--P and GRF residuals are defined relative to different smooth reference fields, both constructions are designed to isolate fluctuations about the appropriate background associated with each method.

Figure~\ref{fig:maps_comparison} compares the residual convergence fields obtained from the S--P and GRF realizations for the same particle mass, $m_{\psi}=10^{-22}\,\mathrm{eV}$. The two approaches produce broadly comparable characteristic fluctuation scales and fractional-residual morphologies. This agreement in their projected fluctuation statistics supports the GRF treatment as an efficient statistical approximation when a detailed description of the underlying wave dynamics is not required and moderate predictive precision is sufficient.

\subsection{GRF Realizations Constructed from the S--P Density Field}
\label{app:sp_grf_construction}

We follow the GRF prescription of \citet{Amruth2023}, but construct both the smooth component and the fluctuation amplitude directly from our S--P simulations rather than from an analytic NFW profile.

For the selected image-aligned S--P convergence map, we first determine the projected center, axis ratio, and position angle from its convergence-weighted second moments. We then construct concentric, self-similar elliptical annuli and define the smooth convergence field as
\begin{equation}
    \kappa_{\rm sm}(\boldsymbol{\xi})
    \equiv
    \left\langle
        \kappa_{\rm SP}(\boldsymbol{\xi})
    \right\rangle_{\rm ell},
    \label{eq:grf_elliptical_smooth}
\end{equation}
where $\boldsymbol{\xi}$ denotes the position in the lens plane and $\langle\cdots\rangle_{\rm ell}$ denotes the mean within the corresponding elliptical annulus. See Section~\ref{app:residual_maps} for further discussion.

The spatially varying variance of the convergence fluctuations is evaluated directly from the corresponding three-dimensional S--P density field using the line-of-sight integral in Equation~(8) of
\citet{Amruth2023},
\begin{equation}
    \sigma_{\kappa}^{2}(\boldsymbol{\xi})
    =
    \frac{\lambda_{\rm dB}}
         {\Sigma_{\rm cr}^{2}}
    \int
        \rho_{\rm SP}^{2}(z,\boldsymbol{\xi})
    \,{\rm d}z ,
    \label{eq:grf_variance_sp}
\end{equation}
thereby avoiding the analytic expression specific to an NFW profile. Before performing the integration, the density field is rotated to the same viewing direction as that used for the corresponding S--P projection. For a uniform cubic grid with cell width $\Delta$ and voxel masses $m_{ijk}$, Equation~(\ref{eq:grf_variance_sp}) is implemented as
\begin{equation}
    \sigma_{\kappa,ij}^{2}
    =
    \frac{\lambda_{\rm dB}}
         {\Sigma_{\rm cr}^{2}\Delta^{5}}
    \sum_{k} m_{ijk}^{2}.
    \label{eq:grf_variance_discrete}
\end{equation}
The resulting variance map is then transformed using the same two-dimensional orthogonal transformation and translation applied to the corresponding image-aligned convergence map. See Section~\ref{app:method} for further discussion.

The GRF convergence fluctuation, $\delta\kappa_{\rm GRF}$, is generated following \citet{Amruth2023}, with its local amplitude set by
$\sigma_{\kappa}(\boldsymbol{\xi})$. The final S--P-constructed GRF realization is therefore
\begin{equation}
    \kappa_{\rm FDM}^{\rm GRF}(\boldsymbol{\xi})
    =
    \kappa_{\rm sm}(\boldsymbol{\xi})
    +
    \delta\kappa_{\rm GRF}(\boldsymbol{\xi}).
    \label{eq:grf_total_sp}
\end{equation}
Thus, both the smooth convergence profile and the spatial variation of the fluctuation amplitude are constructed from the selected S--P realization.

We select one representative S--P realization to define $\kappa_{\rm sm}$ and $\sigma_{\kappa}$, and generate 1000 independent GRF realizations from this common baseline, thereby obtaining 1000 corresponding convergence maps. The resulting representative residual maps and position-anomaly distributions are shown in Figures~\ref{fig:maps_comparison}(c) and \ref{fig:fdmgrf}. Although this S--P-constructed GRF still does not reproduce the observed image positions as well as the S--P simulations, it provides a noticeable improvement for radio2. In particular, the median position anomaly shifts from $\sim 16\,{\rm mas}$ to $\sim12\,{\rm mas}$, while the smallest anomaly among the best-fitting realizations decreases from $\sim 7\,{\rm mas}$ to $\sim 6\,{\rm mas}$. This reduction in the median position anomaly is interesting as this may indicate that the smooth S--P halo used as the underlying global mass distribution for the construction of the GRF is indeed a more accurate description for the global halo of the lensing galaxy. In addition, the fractional-residual maps no longer exhibit the pronounced central concentration found when an analytic NFW profile is adopted as the smooth component (which makes sense as we do not include a central soliton component in addition to the NFW profile), and instead display a more spatially stochastic fluctuation pattern.
\begin{figure*}[!tbh]
    \centering
    \includegraphics[width=\textwidth]
    {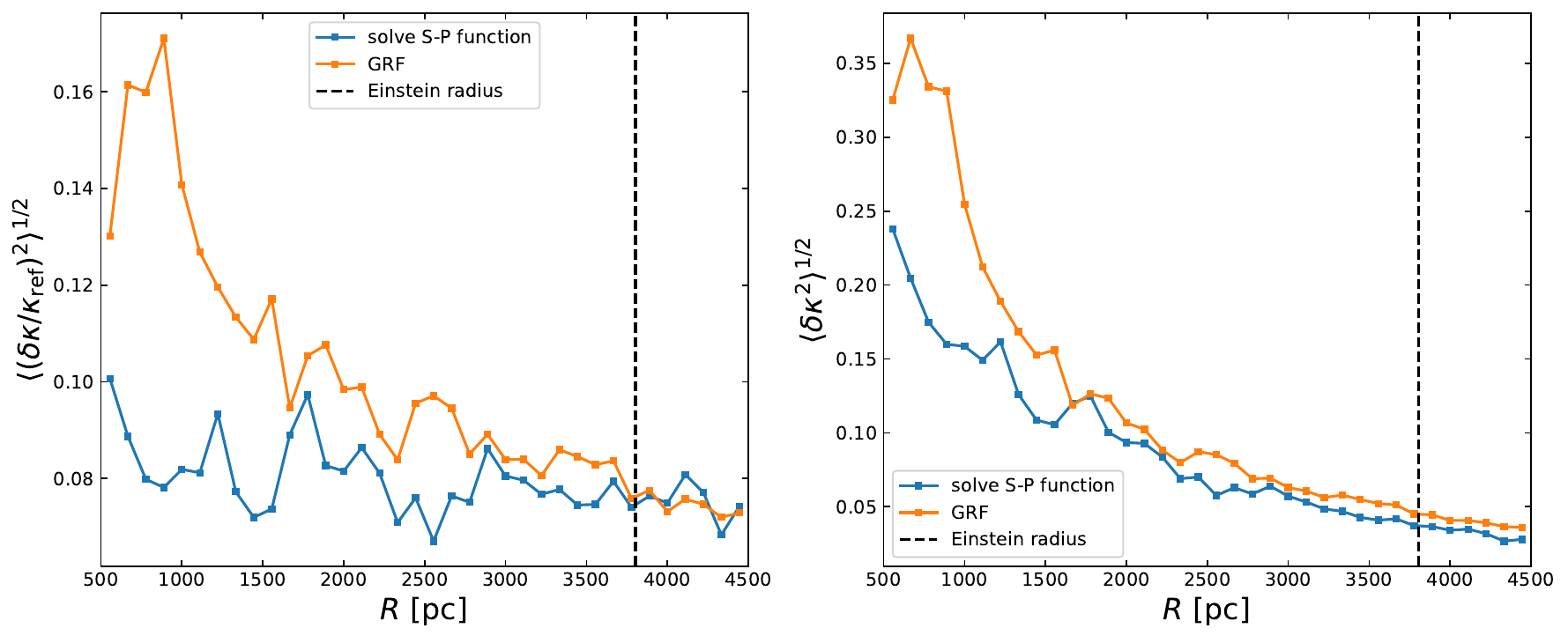}
    \caption{
    Annular rms amplitudes of the projected convergence fluctuations for the Schr\"odinger--Poisson (S--P) and Gaussian random field (GRF) realizations. The left and right panels show the fractional rms, $\langle(\delta\kappa/\kappa_{\rm ref})^2\rangle^{1/2}$, and the absolute rms, $\langle(\delta\kappa)^2\rangle^{1/2}$, respectively. Here, $\delta\kappa=\kappa-\kappa_{\rm ref}$, with $\kappa_{\rm ref}=\langle\kappa_{\rm FDM}\rangle_{\rm ell}$ for the S--P realization and $\kappa_{\rm ref}=\kappa_{\rm NFW}$ for the GRF realization. The vertical dashed lines mark the effective Einstein radius. See Section~\ref{app:comparison} for further discussion.
    }
    \label{fig:statistics}
\end{figure*}

\subsection{Origin of the Improved S--P Lensing Predictions Relative to the GRF Approximation}
\label{app:sp_grf_lensing_difference}

The preceding comparisons show that the Schr\"odinger--Poisson (S--P) and Gaussian random field (GRF) approaches produce broadly similar projected fluctuation statistics, while their lensing predictions are not identical. In particular, the S--P realizations provide better agreement with the observed image positions of HS~0810+2554 than the conventional GRF realizations considered in this work. Here we briefly discuss why these two results are not contradictory and which additional physical information is retained by the direct S--P treatment.

\paragraph{Factors leading to different lensing predictions despite similar fluctuation statistics.}
First, the quantities compared in Figure~\ref{fig:statistics} are annular statistics. They characterize the average fluctuation amplitude at a given elliptical radius, but do not retain the full local spatial structure of an individual realization. Such local structure is particularly relevant for strong lensing, where the position of an individual image responds to the mass distribution in its immediate neighborhood.

More importantly, the lensing calculation depends on the \emph{total} projected mass distribution rather than on the fluctuation field alone. For the S--P calculation, the total three-dimensional density field is obtained directly from the self-consistent wave evolution. The decomposition into a smooth component and a fluctuation component introduced in Section~\ref{app:residual_maps} is therefore only a diagnostic used to compare with the GRF prescription. By contrast, the conventional GRF construction begins with an assumed smooth halo, determines the fluctuation amplitude relative to this background, and then superposes the generated fluctuations on it. Similar fluctuation statistics therefore do not imply identical smooth components, total mass distributions, or lensing predictions.

\begin{figure*}[t]
    \centering

    \includegraphics[
        width=\textwidth,
        height=0.22\textheight,
        keepaspectratio
    ]{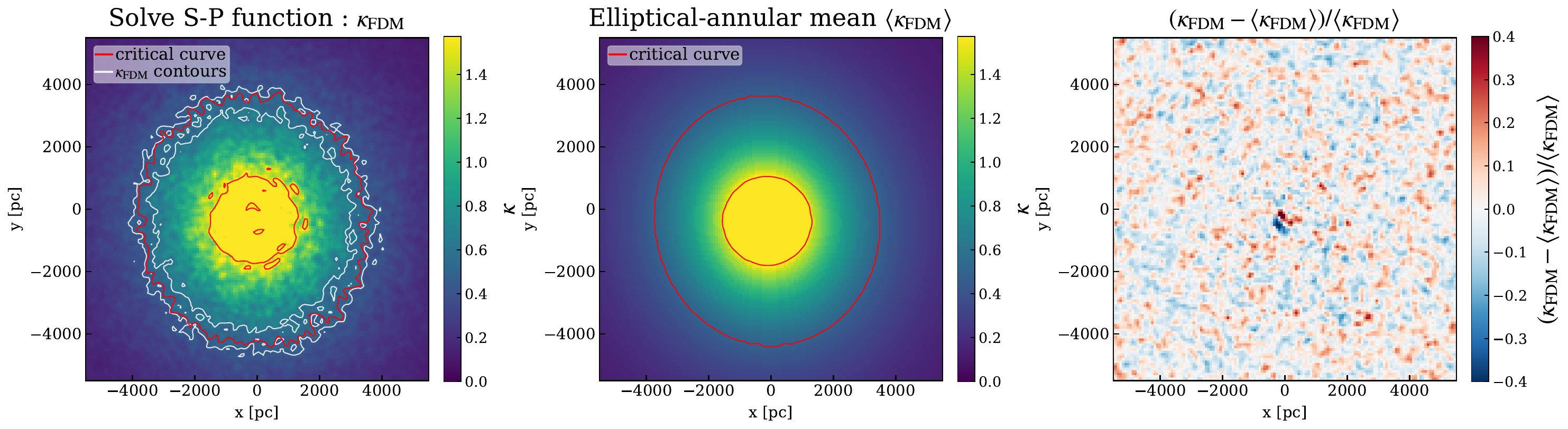}

    \par\vspace{0.05cm}
    \textbf{(a) S--P realization}

    \vspace{0.15cm}

    \includegraphics[
        width=\textwidth,
        height=0.22\textheight,
        keepaspectratio
    ]{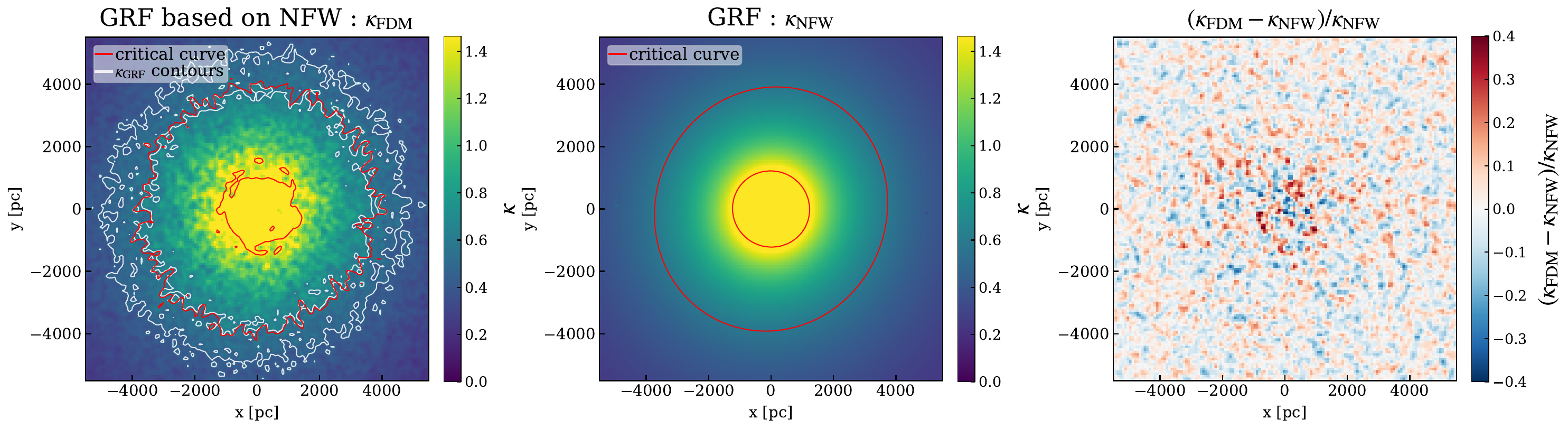}

    \par\vspace{0.05cm}
    \textbf{(b) GRF realization based on NFW}

    \vspace{0.15cm}

    \includegraphics[
        width=\textwidth,
        height=0.22\textheight,
        keepaspectratio
    ]{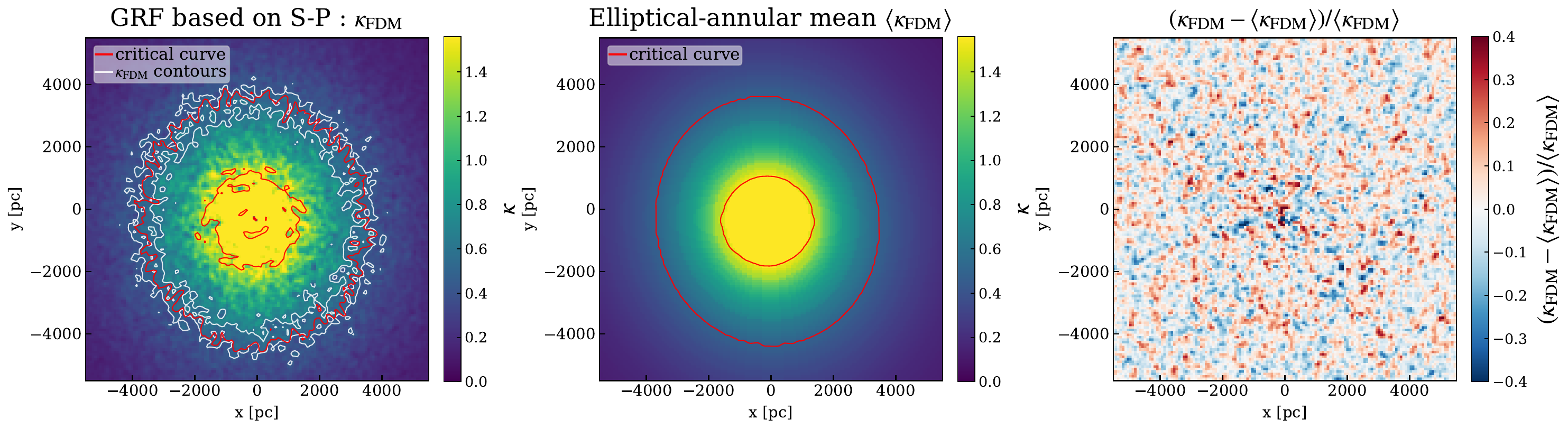}

    \par\vspace{0.05cm}
    \textbf{(c) GRF realization based on the S--P realization}

    \caption{
    Extension of Figure~\ref{fig:map_fdm}; see Section~\ref{app:residual_maps} and Section~\ref{app:sp_grf_construction} for further discussion. Panel set (a) shows the $m_{\psi}=10^{-22}\,{\rm eV}$ Schr\"odinger--Poisson (S--P) realization: the projected convergence field $\kappa_{\rm FDM}$, its elliptical-annular mean $\langle\kappa_{\rm FDM}\rangle$, and the corresponding fractional residual $(\kappa_{\rm FDM}-\langle\kappa_{\rm FDM}\rangle)/ \langle\kappa_{\rm FDM}\rangle$. Panel set (b) shows the conventional Gaussian random field (GRF) realization based on the NFW model for $\lambda_{\rm dB}\sim 120\,{\rm pc}$: the total convergence field $\kappa_{\rm FDM}$, the smooth NFW reference field $\kappa_{\rm NFW}$, and the fractional residual $(\kappa_{\rm FDM}-\kappa_{\rm NFW})/\kappa_{\rm NFW}$. Panel set (c) shows a GRF realization constructed from the S--P result, for which the elliptical-annular mean of the selected S--P convergence map defines the smooth component and the spatially varying fluctuation amplitude is derived from the corresponding three-dimensional S--P density field. The three columns show the resulting total convergence field $\kappa_{\rm FDM}$, its S--P-derived elliptical-annular mean $\langle\kappa_{\rm FDM}\rangle$, and the corresponding fractional residual $(\kappa_{\rm FDM}-\langle\kappa_{\rm FDM}\rangle)/ \langle\kappa_{\rm FDM}\rangle$, respectively. The white contours correspond to $\kappa_{\rm FDM}\sim0.45$ and $0.60$, respectively, and highlight the density fluctuations around the critical curve, which is shown in red. For visualization, each convergence map is capped at its 95th-percentile value, with all values above this threshold assigned the threshold value. Across all three panel sets, the fractional-residual maps are displayed using a common symmetric color scale whose limits are set by the smallest of their individual maximum absolute displayed values, thereby facilitating direct comparison.
    }
    \label{fig:maps_comparison}
\end{figure*}

\begin{figure*}[t]
    \centering \includegraphics[width=0.98\textwidth]{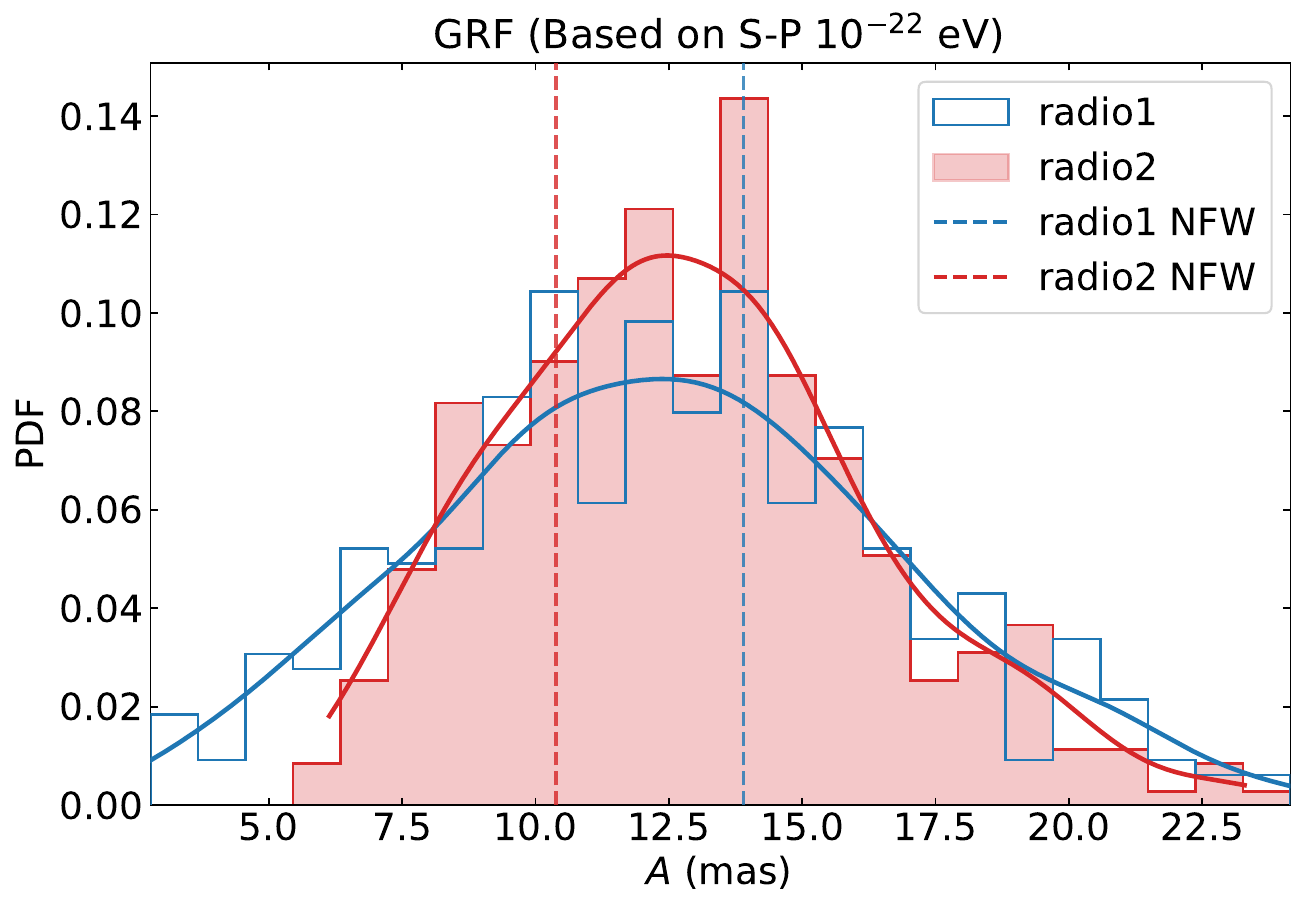}
    \caption{
    Extension of Figure~\ref{fig:Ar}; see Section~\ref{app:sp_grf_construction} for further discussion. Probability distributions of the image-position anomaly $A$ for GRF-based FDM realizations constructed using the selected Schr\"odinger--Poisson (S--P) density field with $m_{\psi}=10^{-22}\,{\rm eV}$. The blue and red histograms show the normalized distributions for radio1 and radio2, respectively, while the solid curves denote the corresponding kernel-density-estimation (KDE) fits. The blue and red vertical dashed lines indicate the position anomalies predicted by the best-fitting smooth NFW model for radio1 and radio2, respectively.
    }
    \label{fig:fdmgrf}
\end{figure*}

This distinction is directly illustrated by Figures~\ref{fig:maps_comparison} and \ref{fig:fdmgrf}. In the GRF approach, the fluctuation amplitude generally traces the smooth halo decreases with radius. When the smooth component and fluctuation amplitude of the GRF are instead constructed from the S--P density field, as described in Section~\ref{app:sp_grf_construction}, the resulting GRF lensing prediction improves, particularly for radio2. This provides a useful example that the underlying large-scale mass distribution and its modulation of the fluctuations are important for the final lensing signal. Figure~\ref{fig:rho} likewise shows that the dynamically evolved S--P profile is only approximately NFW-like in the outer halo and also contains a solitonic central component. Therefore, a GRF approach based on an analytic NFW profile need not reproduce the full structure of the S--P-evolved halo, potentially leading to different lensing predictions.

\paragraph{Additional physical information retained by the S--P evolution.}
The direct S--P calculation also improves upon the GRF description by removing several statistical assumptions. In the conventional GRF approach, the fluctuations are generated as a Gaussian random field whose amplitude is modulated by the adopted smooth density profile. Consequently, their amplitude generally follows the radial variation of that profile. The self-consistent S--P evolution, however, naturally includes spatial correlations, mode coupling, and weak non-Gaussian higher-order structure, which can produce local departures from such a simple modulation.

\begin{figure*}[t]
    \centering \includegraphics[width=0.98\textwidth]{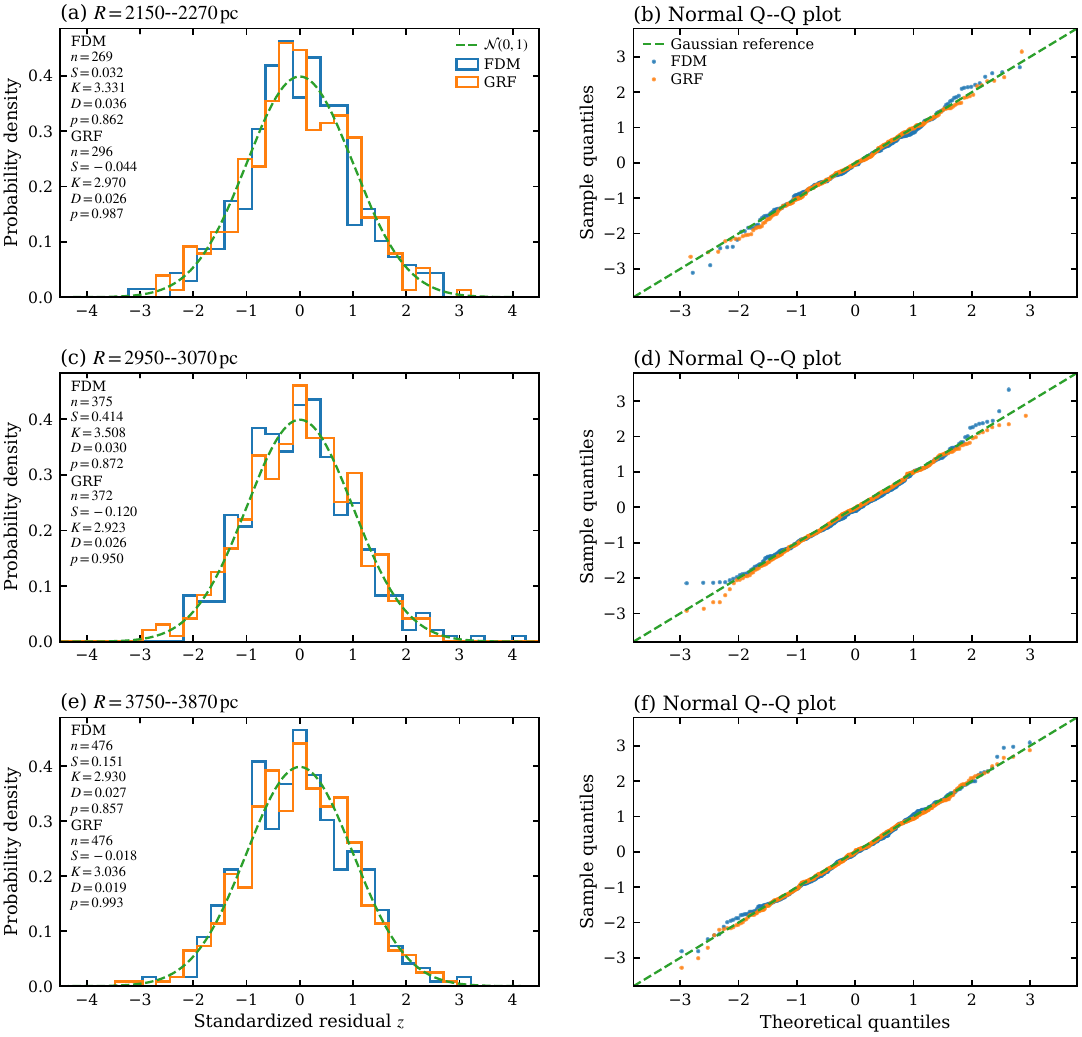}
    \caption{
    Gaussianity tests for the standardized residual fluctuations in three representative annuli. To match the adopted de Broglie wavelength, $\lambda_{\rm dB}\sim 120\,{\rm pc}$, each annulus is assigned a radial width of $\Delta R=120\,{\rm pc}$. For each annulus, the left panel compares the probability-density distributions of the standardized residuals from the fuzzy dark matter (FDM) and Gaussian random field (GRF) realizations with the standard normal distribution, $\mathcal{N}(0,1)$, while the right panel shows the corresponding normal quantile--quantile (Q--Q) plot. The reported statistics are the sample size $n$, skewness $S$, kurtosis $K$, Kolmogorov--Smirnov (K--S) statistic $D$, and its associated $p$-value. The Schr\"odinger--Poisson FDM residuals are broadly consistent with Gaussian statistics within the selected annuli and exhibit good agreement with the GRF realization. See Section~\ref{app:comparison} for further discussion.
    }
    \label{fig:gaussianity}
\end{figure*}

The Gaussianity tests in Figure~\ref{fig:gaussianity} show that the S--P fluctuations are indeed broadly consistent with Gaussian statistics at fixed radius, which explains the overall success of the GRF approximation. However, small departures remain (may be the small non-Gaussian signatures), particularly in the tails of the distributions. These deviations are not fully characterized by the variance and two-point statistics used to construct the GRF and may become relevant for high-precision lensing predictions. A more complete investigation of this higher-order information, including three- and four-point correlation functions and the bispectrum and trispectrum, is left for future work.

A related limitation arises from the central-limit theorem underlying the projected GRF approximation. The projection of a finite halo contains only a finite number of coherence elements. As a simple illustration, a $\sim30\,{\rm kpc}$ line of sight containing structures with a characteristic scale of $\sim200\,{\rm pc}$ samples only $\mathcal{O}(10^2)$, or approximately 150, characteristic coherence lengths. The ideal Gaussian limit therefore need not be reached exactly. Direct S--P evolution avoids this assumption by explicitly evolving the three-dimensional wave-interference field before projection.

Taken together, these results explain both the similarity and the remaining differences between the two approaches. The GRF prescription successfully captures the leading fluctuation scale and low-order projected statistics and therefore provides an efficient approximation when moderate predictive precision is sufficient. The S--P approach, however, self-consistently determines the total density field while retaining the spatial correlations, mode coupling, and higher-order structure generated by the wave dynamics. These additional ingredients naturally propagate into the detailed projected mass distribution and provide a plausible physical origin for the improved image-position predictions obtained from the S--P realizations.

\clearpage
\bibliographystyle{aasjournalv7}
\bibliography{refs}

\end{document}